\def\fr#1#2{\hbox{${#1\over #2}$}}

\def\ni{\noindent}
\def\vs{\vskip.3cm}

\def\+{{(+)}}  \def\-{ {(-)} }   \def\0{ {(0)} }
\def\1{ {(1)} }  \def\2{ {(2)} }

\def\sq{Q\kern-6pt/}
\def\sQ{Q\kern-12pt\nearrow}
\documentclass[11pt,eqs]{article}

\usepackage{latexsym}
\usepackage{amssymb}

\def\be{\begin{equation}}             \def\ee{\end{equation}}
\def\ba{\begin{array}{rcl}}           \def\ea{\end{array}}
\def\beqa{\begin{eqnarray} }          \def\eeqa{\end{eqnarray} }
\def\beqalign{\begin{eqalign}}        \def\eeqalign{\end{eqalign}}
             
\def\bsubeq{\begin{subequations}}     \def\esubeq{\end{subequations}}
\def\bitem{\begin{itemize}}           \def\eitem{\end{itemize}}

\def\DJ{\leavevmode\setbox0=\hbox{D}\kern0pt
 \rlap{\kern.04em\raise.188\ht0\hbox{-}}D}
\def\dj{\leavevmode\setbox0=\hbox{d}\kern0pt
 \rlap{\kern.215em\raise.46\ht0\hbox{-}}d}

\newcommand{\bd}{\begin{displaymath}}
\newcommand{\ed}{\end{displaymath}}
\begin{document}

\title{ Gauge symmetries decrease the number of Dp-brane dimensions. II. Inclusion of the Liouville term
\thanks{Work supported in part by the Serbian Ministry of Science and
Environmental Protection, under contract No. 141036.}}
\author{B. Nikoli\'c \thanks{e-mail address: bnikolic@phy.bg.ac.yu} and B. Sazdovi\'c
\thanks{e-mail address: sazdovic@phy.bg.ac.yu}\\
       {\it Institute of Physics, 11001 Belgrade, P.O.Box 57, Serbia}}
\maketitle
\begin{abstract}

The presence of the antisymmetric background field $B_{\mu\nu}$ leads to the
noncommutativity of the Dp-brane manifold, while the linear dilaton field in
the form $\Phi(x)=\Phi_0+a_\mu x^\mu$, causes the appearance of the
commutative Dp-brane coordinate, $x_c=a_\mu x^\mu$. In the present article we consider the case where the conformal invariance is realized by inclusion of the Liouville term. Then, the theory is conformally invariant even in the presence of the world sheet conformal factor $F$, and it depends on the new parameter, the central charge $c$. As well as in
the absence of the Liouville action, for particular relations between background fields, the local gauge symmetries appear in the
theory. They turn some Neumann
boundary conditions into the Dirichlet ones, and decrease the number of the
Dp-brane dimensions.

\end{abstract}
\vs

\ni {\it PACS number(s)\/}: 11.10.Nx, 04.20.Fy, 11.25.-w   \par

\section{Introduction}

When the ends of the open string are attached to a Dp-brane with antisymmetric Kalb-Ramond field $B_{\mu \nu}$, the Dp-brane world-volume becomes noncommutative \cite{CDS}. The presence of the linear dilaton field $\Phi$, \cite{LRR}-\cite{PR}, turns one Dp-brane coordinate, $x_c=x^\mu\partial_\mu \Phi$, to commutative one and the conformal part of the world-sheet metric, $F$, to additional noncommutative variable \cite{BS2}. 

The possible breaking of the conformal invariance in the open string theory by the boundary conditions has been investigated in Ref.\cite{PR}. It was shown that, besides vanishing of the $\beta$ functions, in the case of the linear dilaton there are additional conditions that dilaton gradient $a_i=\partial_i \Phi$ must satisfy. It should be lightlike vector, either with respect to the closed string metric, $a^2\equiv G^{ij}a_ia_j=0$, or with respect to the open string (effective) metric, $\tilde a^2\equiv(G_{eff}^{-1})^{ij}a_i a_j=0$. The above restrictions decrease the number of the Dp-brane dimensions, turning some Neumann to Dirichlet boundary conditions.

In the present paper we change the conditions for quantum conformal invariance. The usual requirement is vanishing of all three
$\beta$ functions corresponding to background fields, $\beta^G_{\mu\nu}=\beta^B_{\mu\nu}=\beta^\Phi=0$. Here, we use the fact that the vanishing of two
$\beta$ functions, corresponding to the metric and antisymmetric field, $\beta^G_{\mu\nu}=\beta^B_{\mu\nu}=0$,
implies that the third one, corresponding to dilaton field, is constant, $\beta^\Phi=c$, \cite{FCB}.
Instead to choose this constant to be zero, as we did in the previous paper \cite{PR}, in this article we add Liouville term in order to cancel constant contribution to the conformal anomaly. This approach is more general because the theory and, particulary, the noncommutativity parameter depend on arbitrary central charge $c$. The advantage is achievement of the conformal invariance without requirement for decoupling of the conformal factor of the world-sheet metric, $F$. Consequently, for $c\neq0$ the presence of the field $F$ in boundary conditions does not break conformal invariance as in Ref.\cite{PR}.

In order to clarify notation and terminology we will distinguish two descriptions of the open string theory. We start with variable $x^\mu$ and background field $G_{\mu\nu}$, where the theory is described by equations of motion and boundary conditions. We are able to solve boundary conditions and introduce the effective theory defined only by equations of motion. This is again the string theory, but in terms of effective coordinates $q^\mu$ (symmetric
under transformation $\sigma \to -\sigma$) and effective background field $G_{\mu\nu}^{eff}$. Following Seiberg and Witten \cite{CDS}, we use the names closed string metric for $G_{\mu\nu}$ and open string metric for $G_{\mu\nu}^{eff}$ (the metric tensors seen by the closed and open string, respectively).

The Liouville action itself is the kinetic term for the field $F$. So, we are going to treat it equally with other variables. In particular we choose Neumann boundary condition for $F$. Note that, although by simple changes of variables the new field ${}^\star F$
decouples, and the term with linear dilaton disappears, the case is nontrivial because the
new metric ${}^\star G_{ij}$ and the corresponding effective one ${}^\star G_{ij}^{eff}$ become singular for $\alpha a^2=1$ and
$\alpha \tilde a^2=1$, respectively ($\alpha$ is an useful constant defined in (\ref{eq:alfa}) proportional to the inverse central charge). We use the mark star ($\star$) to distinguish description in terms of variables ($x^i\, ,{}^\star F$) from that of ($x^i\, ,F$).

Up to the changing the conditions on the dilaton gradient $a_i$ ($a^2=0 \to  a^2=\frac{1}{\alpha}$ and $\tilde a^2=0 \to \tilde a^2=\frac{1}{\alpha}$), there is a complete analogy of the noncommutativity properties with
the cases of the previous paper \cite{PR}. Note that, here we have whole one-parameter class of theories with the same properties, and in particular, the result of the previous paper has been obtained for ($\alpha\to\infty \Leftrightarrow c=0$).

In the first case, $1-\alpha a^2$
is a coefficient in front of the velocity $\dot x_0=a_i\dot x^i$, so that condition $\alpha a^2=1$ produces the
standard canonical constraint. By simple analysis we conclude that it is of the first class. For $\alpha \tilde a^2=1$, some of the constraints originating from the boundary conditions, turn from the second class into the first
class constraints. 

The first class constraints generate local gauge symmetries. They turn some of the initial Neumann boundary conditions into Dirichlet ones and
decrease the number of the Dp-brane dimensions. The string coordinates, which depend on the effective ones but also on the
corresponding momenta, define the noncommutative subspace of Dp brane. The noncommutativity
parameter is proportional to the antisymmetric field $B_{ij}$. The field ${}^\star F$ decouples from the rest.  So, it plays the role of the commutative variable instead the variable $x_c=a_i x^i$ in the case without Liouville term.

At the end of paper, in Concluding remarks, we summarize the results of the investigation. Also there are three appendices. The first one is devoted to the projectors, which help us to express the results clearly. In the second appendix we introduce the redefined closed and open string star metrics, while in the third one we discussed the separation of the center of mass variables.

\section{Conformal invariance with the help of Liouville action}
\setcounter{equation}{0}

The action  
\begin{equation}\label{eq:action2}
S_{(G+B+\Phi)} = \kappa  \int_\Sigma  d^2 \xi  \sqrt{-g}  \left\{  \left[  {1
\over 2}g^{\alpha\beta}G_{\mu\nu}(x) +{\varepsilon^{\alpha\beta}
\over \sqrt{-g}}  B_{\mu\nu}(x) \right] \partial_\alpha x^\mu
\partial_\beta x^\nu +  \Phi (x) R^{(2)}  \right\} \,  ,
\end{equation}
describes the evolution of the open string
in the background consisting of the space-time metric
$G_{\mu\nu}(x)$, Kalb-Ramond antisymmetric field $B_{\mu\nu}(x)$,
and dilaton scalar field $\Phi(x)$ (for more details see \cite{P}). The world sheet $\Sigma$ is
parameterized by $\xi^\alpha=(\tau\, ,\sigma)$ ($\alpha=0,1$), and
the $D$-dimensional space-time by the coordinates $x^\mu$
($\mu=0,1,2,\dots,D-1$). The intrinsic world sheet metric is
$g_{\alpha\beta}$ and $R^{(2)}$ is the related scalar curvature.

There are three $\beta$ functions corresponding to the space-time
metric $G_{\mu\nu}$, antisymmetric field $B_{\mu\nu}$, and dilaton
field $\Phi$
\begin{equation}\label{eq:betaG}
\beta^G_{\mu \nu} \equiv  R_{\mu \nu} - \frac{1}{4} B_{\mu \rho
\sigma} B_{\nu}{}^{\rho \sigma} +2 D_\mu a_\nu    \,  ,
\end{equation}
\begin{equation}\label{eq:betaB}
\beta^B_{\mu \nu} \equiv  D_\rho B^\rho{}_{\mu \nu} -2 a_\rho
 B^\rho{}_{\mu \nu}  \,  ,
\end{equation}
\begin{equation}\label{eq:betaFi}
\beta^\Phi \equiv 2\pi \kappa{D-26 \over 6} - \fr{1}{24} B_{\mu \rho \sigma}
B^{\mu \rho \sigma} -  D_\mu a^\mu + 4 a^2  \,  ,
\end{equation}
which characterize the conformal anomaly of the sigma model
(\ref{eq:action2}). The space-time Ricci tensor and covariant
derivative are denoted with $R_{\mu \nu}$ and $D_\mu$,
respectively, while $B_{\mu \rho \sigma}=\partial_\mu
B_{\nu\rho}+\partial_\nu B_{\rho\mu}+\partial_\rho B_{\mu\nu}$ is
the field strength for the field $B_{\mu\nu}$ and
$a_\mu=\partial_\mu \Phi$ is the gradient of the dilaton field.

It is known from Ref.\cite{FCB} that vanishing of $\beta^G_{\mu
\nu}$ and $\beta^B_{\mu \nu}$ implies constant value of the third
$\beta$ function, $\beta^\Phi=c$. We choose particular solution of the Eqs.(\ref{eq:betaG}) and (\ref{eq:betaB})
\begin{equation}
G_{\mu \nu}(x) = G_{\mu \nu}=const \,  , \quad B_{\mu \nu} (x) =
B_{\mu \nu} = const\, ,\quad 
\Phi(x) =\Phi_0 + a_\mu x^\mu  \,  . \quad (a_\mu = const)
\end{equation}
Then Eq.(\ref{eq:betaFi}) produces the condition
\begin{equation}\label{eq:betaFis}
\beta^\Phi=2\pi \kappa{D-26 \over 6}+4 a^2\equiv c\, ,
\end{equation}
under which the above solution is consistent with all equations of motion. On these conditions, the
non-linear sigma model (\ref{eq:action2}) becomes conformal field
theory. There exists a Virasoro algebra with central charge $c$.

The remaining anomaly, represented by the Schwinger term of the
Virasoro algebra, can also be cancelled by introducing
corresponding Wess-Zumino term, which in the case of the conformal
anomaly takes the form of the Liouville action
\begin{equation}\label{eq:kinclan}
S_{L}=  -\frac{\beta^\Phi}{2(4\pi)^2\kappa}\int_{\Sigma} d^2 \xi \sqrt{-g}
R^{(2)}\frac{1}{\Delta}R^{(2)}\, ,\quad
\Delta=g^{\alpha\beta}\nabla_\alpha\partial_\beta\, ,
\end{equation}
where $\nabla_\alpha$ is the covariant derivative with respect to
the intrinsic metric $g_{\alpha\beta}$. Appropriate choice of the
coefficient in front of Liouville action makes the theory fully conformally invariant and the complete action takes
the form
\begin{equation}\label{eq:2deo1}
S=S_{(G+B+\Phi)}+S_L \,  .
\end{equation}
We choose a particular background, decomposing the space-time coordinates $x^\mu(\xi)$ in Dp-brane
coordinates denoted by $x^i(\xi) \,  (i =0,1,...,p)$ and the
orthogonal ones $x^a(\xi) \,  (a = p+1,p+2,...,D-1)$, in such a way that $G_{\mu \nu} = 0$, ($\mu =i  \,  ,\nu = a$). For the other two background fields we choose: $B_{\mu \nu} \to
B_{i j}$, $a_\mu \to a_i$ i.e. they are nontrivial only on the
Dp brane. The part of the action describing the string
oscillation in $x^a$ directions decouples from the rest.

Imposing the conformal gauge
$g_{\alpha\beta}=e^{2F}\eta_{\alpha\beta}$, we obtain $R^{(2)}=-2\Delta F$ and the action (\ref{eq:2deo1}) takes the form
\begin{equation}\label{eq:2deo2}
S=\kappa \int_{\Sigma}d^2 \xi\bigg [\bigg
(\frac{1}{2}\eta^{\alpha\beta}G_{ij}+\epsilon^{\alpha\beta} B_{ij}
\bigg)\partial_{\alpha}x^{i}\partial_{\beta}x^{j}+2\eta^{\alpha\beta}a_i\partial_\alpha x^i \partial_\beta F+\frac{2}{\alpha}\eta^{\alpha\beta}
\partial_{\alpha}F \partial_{\beta}F\bigg] \,  ,
\end{equation}
where we introduce useful notation
\begin{equation}\label{eq:alfa}
\frac{1}{\alpha}=\frac{\beta^\Phi}{(4\pi\kappa)^2} \, .
\end{equation}
The field $F$ is a dynamical variable with the Liouville action as a kinetic term. In order to cancel the term linear in $F$, we change the variables, $F\to {}^\star
F=F+\frac{\alpha}{2}a_i x^i$, and obtain
\begin{equation}\label{eq:2deo}
S=\kappa \int_{\Sigma}d^2 \xi \bigg [\bigg
(\frac{1}{2}\eta^{\alpha\beta}\;{}^\star
G_{ij}+\epsilon^{\alpha\beta} B_{ij}
\bigg)\partial_{\alpha}x^{i}\partial_{\beta}x^{j}+\frac{2}{\alpha}\eta^{\alpha\beta}
\partial_{\alpha}{}^\star F\partial_{\beta}{}^\star F\bigg] \,  .
\end{equation}
This is a standard form of the action without dilaton term and with redefined Liouville term, $F \to {}^\star F$, and redefined space-time metric, ${}^\star G_{ij}=G_{ij}-\alpha a_i a_j$. The dilaton dependence is now through the metric ${}^\star G_{ij}$. 

We choose Neumann boundary conditions for the redefined conformal factor of the intrinsic metric ${}^\star F$. The field ${}^\star F$ completely decouples, as well as the coordinate $x^a$, but because of its Neumann boundary conditions, we will treat it as a Dp-brane variable. In all cases it is a commutative variable.

All nontrivial features of the model (\ref{eq:2deo}) follow from the fact that the star metrics (${}^\star
G_{i j}$ and the corresponding effective one ${}^\star G_{ij}^{eff}$) are singular and consequently they produce gauge symmetries of the
theory. It is easy to check that for $\alpha a^2=1$ and $\alpha \tilde a^2=1$ we have $\det {}^\star G_{ij}=0$ and $\det {}^\star G_{ij}^{eff}=0$, respectively.

\section{Noncommutativity for regular star metrics ${}^\star G_{ij}$ and ${}^\star G^{eff}_{ij}$ ($\alpha a^2\neq1$ and $\alpha \tilde a^2\neq1$)}
\setcounter{equation}{0}

In this section we will analyze the case when both the metric ${}^\star G_{ij}$ and the corresponding effective one ${}^\star G_{ij}^{eff}$ are nonsingular. Up to the field ${}^\star F$, which is decoupled from the other Dp-brane variables, there is complete formal analogy with the case without dilaton field with substitution $G_{ij}\to {}^\star G_{ij}$. For comletness we present the main steps of the procedure and add the parts corresponding to ${}^\star F$.

\subsection{Canonical Hamiltonian in terms of currents}

The momenta canonically conjugated to the fields $x^i$ and ${}^\star F$ are
\begin{equation}\label{eq:kanimp}
\pi_i=\kappa ({}^\star G_{ij}\dot x^j-2B_{ij}x'^j)\, ,\quad \pi=\frac{4\kappa}{\alpha}{}^\star \dot F\, .
\end{equation}
Using the definition of the canonical Hamiltonian $\mathcal
H_c=\pi_i \dot x^i+\pi \;{}^\star \dot F-\mathcal L$, we obtain
\begin{eqnarray}
H_c&=&\int d\sigma \mathcal H_c\, ,\qquad \mathcal
H_{c}=T_{-}-T_{+}\, , \nonumber \\
T_{\pm}&=&\mp\frac{1}{4\kappa}\left[ ({}^\star
G^{-1})^{ij}\;{}^\star j_{\pm i} \;{}^\star j_{\pm
j}+\frac{\alpha}{4} \;{}^\star j_{\pm(F)} \;{}^\star
j_{\pm(F)}\right] \label{eq:hamilton}\, ,
\end{eqnarray}
where
\begin{equation}\label{eq:zvezdastruje}
{}^\star j_{\pm i}=\pi_i+2\kappa\;{}^\star \Pi_{\pm ij}x'^j\, ,\quad
{}^\star j_{\pm(F)}=\pi\pm\frac{4\kappa}{\alpha}{}^\star F'\,
,\quad \left( {}^\star \Pi_{\pm ij}=B_{ij}\pm\frac{1}{2}{}^\star
G_{ij}\right)
\end{equation}
and the inverse metric $({}^\star G^{-1})^{ij}$ is introduced in
Eq.(\ref{eq:Lmetrikain}).

From the basic Poisson bracket algebra
\begin{equation}\label{eq:azs}
\left\lbrace x^i(\tau,\sigma),\pi_j(\tau,\overline\sigma)\right\rbrace=\delta^i{}_j\delta(\sigma-\overline\sigma)\, ,\quad \left\lbrace  {}^\star F(\tau,\sigma),\pi(\tau,\overline\sigma)\right\rbrace=\delta(\sigma-\overline\sigma)\, ,
\end{equation}
directly follows the current algebra
\begin{equation}\label{eq:algcur}
\left\lbrace {}^\star j_{\pm i},{}^\star j_{\pm j}\right\rbrace=\pm2\kappa\;{}^\star G_{ij}\delta'\, ,\quad \left\lbrace  {}^\star j_{\pm(F)},{}^\star j_{\pm(F)}\right\rbrace=\pm\frac{8\kappa}{\alpha}\delta'\, ,\quad \left\lbrace  {}^\star j_{\pm i}, {}^\star j_{\pm(F)}\right\rbrace=0\, ,
\end{equation}
while all opposite chirality currents commute and for simplicity we define
$\delta'\equiv \partial_\sigma \delta(\sigma-\overline\sigma)$.
Consequently, the Poisson brackets between canonical Hamiltonian
and the currents ${}^\star j_{\pm i}$ and ${}^\star j_{\pm(F)}$
are proportional to their sigma derivatives
\begin{equation}\label{eq:hamstruja}
\left\lbrace  H_c,{}^\star j_{\pm i}\right\rbrace=\mp {}^\star j'_{\pm i}\, ,\quad \left\lbrace H_c,{}^\star j_{\pm(F)}\right\rbrace=\mp {}^\star j'_{\pm(F)}\, .
\end{equation}

\subsection{Boundary conditions as canonical constraints}

We will use Neumann boundary conditions for the fields $x^i$ and
${}^\star F$. The boundary conditions are of the form
$\gamma_{i}^{(0)} \Big |_0^\pi=0$ and $\gamma^{(0)}\Big
|_0^\pi=0$, where
\begin{equation}
\gamma_{i}^{(0)}=\frac{\partial \mathcal L}{\partial(\partial_\sigma x^i)}=\kappa (-{}^\star G_{ij}x'^j+2B_{ij}\dot x^j)\, ,\quad \gamma^{(0)}=\frac{\partial \mathcal L}{\partial(\partial_\sigma {}^\star F)}=-\frac{4\kappa}{\alpha}\;{}^\star F'\, .
\end{equation}
They can be rewritten in terms of the currents
(\ref{eq:zvezdastruje}) as
\begin{equation}\label{eq:gruslov}
\gamma_{i}^{(0)}=({}^\star \Pi_{+}\;{}^\star G^{-1})_i{}^j \;{}^\star j_{- j}+({}^\star \Pi_{-}\;{}^\star G^{-1})_i{}^j \;{}^\star j_{+ j}   \,  ,
\quad \gamma^{(0)}=\frac{1}{2}\left[  {}^\star j_{-(F)}-{}^\star j_{+(F)}\right] \, ,
\end{equation}
and treated as canonical constraints. Examing the consistency of the constraints, with the help of the relations
(\ref{eq:hamstruja}), we obtain an infinite set of constraints.
Using Taylor expansion, we rewrite all the constraints at $\sigma=0$ in
a more compact $\sigma$-dependent form
\begin{eqnarray}\label{eq:gruslov1}
\Gamma_{i}(\sigma)&=&({}^\star \Pi_{+}\;{}^\star G^{-1})_i{}^j \;{}^\star j_{- j}(\sigma)+({}^\star \Pi_{-}\;{}^\star G^{-1})_i{}^j \;{}^\star j_{+ j}(-\sigma) \,  ,\nonumber \\ \Gamma(\sigma)&=&\frac{1}{2}\left[{}^\star j_{-(F)}(\sigma)-{}^\star j_{+(F)}(-\sigma)\right]\, .\label{eq:velikog}
\end{eqnarray}
In the same way, we obtain similar expressions from the contraints at $\sigma=\pi$. From the fact that the differences of the corresponding constraints at $\sigma=0$ and $\sigma=\pi$ are also constraints, we conclude that all positive chirality currents and, consequently, all variables are $2\pi$ periodic functions. Because of this periodicity the constraints at $\sigma=\pi$ can be discarded (for more details see Ref.\cite{BS2}).

We complete the consistency procedure finding the Poisson brackets
\begin{equation}
\left\lbrace H_c,\Gamma_i\right\rbrace=\Gamma_i'\, ,\quad  \left\lbrace H_c,\Gamma\right\rbrace=\Gamma'\, ,
\end{equation}
which means that there are no more constraints in the theory.

The algebra of the constraints $\chi_A=(\Gamma_i,\Gamma)$ has a simple matrix form
\begin{equation}\label{eq:dklasa}
\left\lbrace \chi_A(\sigma), \chi_B(\overline \sigma) \right\rbrace=-\kappa M_{AB}\delta' \, ,\qquad M_{AB}=\left (
\begin{array}{cc}
{}^\star G^{eff}_{ij} & 0\\
0 & \frac{4}{\alpha}
\end{array}\right )\, .
\end{equation}
The space-time component, which we will call the effective or open
string metric, is defined in Eq.(\ref{eq:effm}). The determinant
\begin{equation}\label{eq:detM}
\det M_{AB}=\frac{4}{\alpha} \tilde A \mathcal A \det G^{eff}_{ij}=\frac{4}{\alpha}\frac{\tilde A^2}{A}\det G_{ij}^{eff}\, ,
\end{equation}
is regular for $\tilde A\equiv 1-\alpha \tilde a^2\neq0$ and
$A\equiv 1-\alpha a^2\neq0$, and all constraints are of the second
class. The fields $G_{ij}$ and $B_{ij}$ are
chosen in such a way that $\det G_{ij}^{eff}\neq0$.

\subsection{Solution of the constraint equations}

Let us introduce the common symbol for the coordinates and their canonically conjugated momenta, $C^A=(x^i\, ,{}^\star F\, ,\pi_i\, ,\pi)$. It is useful to define the symmetric and antisymmetric parts in $\sigma$-parity as
\begin{equation}\label{eq:mena1}
O^A(\sigma)=\frac{1}{2}\left[C^A (\sigma)+C^A (-\sigma)\right] \,  ,
\quad \overline{O}^{A}(\sigma)=\frac{1}{2}\left[
C^{A}(\sigma)-C^{A}(-\sigma)\right] \, ,
\end{equation}
where to $O^A=(q^i\, ,{}^\star f\, ,p_i\, ,p)$ we will refer as the effective variables. In terms of these variables, the constraints $\Gamma_{i}(\sigma)$
and $\Gamma(\sigma)$ have the form
\begin{equation}\label{eq:veza1}
\Gamma_i=2(B\; {}^\star G^{-1})_i{}^j p_j-\kappa \; {}^\star G_{ij}^{eff}\overline q'^j+\overline p_i\, ,\quad \Gamma=
\overline p-\frac{4\kappa}{\alpha}\;{}^\star \overline f'\, .
\end{equation}

From
\begin{equation}
\Gamma_{i}(\sigma)=0\, ,\qquad \Gamma(\sigma)=0\, ,
\end{equation}
choosing integration constants $\overline q^i(\sigma=0)=0$ and ${}^\star \overline f(\sigma=0)=0$, we obtain the solution for string variables expressed in terms of
the effective ones
\begin{equation}\label{eq:solution1}
x^i(\sigma)=q^i(\sigma)-2\;{}^\star \Theta^{ij}\int_0^\sigma d\sigma_1 p_j(\sigma_1)\, ,\quad \pi_i=p_i\, ,
\end{equation}
\begin{equation}\label{eq:solution2}
{}^\star F={}^\star f\, ,\quad \pi=p \, .
\end{equation}
Note that as we explained in introduction, the string variables $x^i$ and $\pi_i$ describe the string dynamics before solving constraints originating from boundary conditions, while the effective ones, $q^i$ and $p_i$, describe the string after solving constraints.

The parameter ${}^\star \Theta^{ij}$ is defined as
\begin{equation}\label{eq:tetka1}
{}^\star \Theta^{ij}=-\frac{1}{\kappa}({}^\star G_{eff}^{-1}B\;{}^\star G^{-1})^{ij}=-\frac{1}{\kappa}(G_{eff}^{-1}\check \Pi_T^0 BG^{-1}\check \Pi_T^0)^{ij} \, ,
\end{equation}
where
\begin{equation}
(\check \Pi_T^0)_i{}^j=\delta_i{}^j-(1-\frac{1}{\tilde A})(\Pi_0)_i{}^j\, ,
\end{equation}
and the projector $(\Pi_0)_i{}^j$ is introduced in Eq.(\ref{eq:pitenula}). 

In terms of ${}^\star G_{ij}$, the parameter ${}^\star \Theta^{ij}$ has the same form as the parameter $\Theta^{ij}$ in terms of $G_{ij}$ in the case without dilaton field. Note that in this approach the noncommutativity parameter ${}^\star \Theta^{ij}$ depends on central charge $c$.

\subsection{Effective theory and noncommutativity relations}

Let us introduce the effective currents
\begin{equation}
{}^\star \tilde j_{\pm i}=p_i\pm\kappa\;{}^\star
G_{ij}^{eff}q'^j\, ,\quad {}^\star \tilde
j_{\pm(F)}=p\pm\frac{4\kappa}{\alpha}\;{}^\star f'\, .
\end{equation}
Using the solution (\ref{eq:solution1})
and (\ref{eq:solution2}) we correlate them with currents given in Eq.(\ref{eq:zvezdastruje}) 
\begin{equation}
{}^\star  j_{\pm i}=\pm2({}^\star \Pi_\pm\;{}^\star
G_{eff}^{-1})_i{}^j \;{}^\star \tilde j_{\pm j}\, ,\quad {}^\star
j_{\pm(F)}={}^\star \tilde j_{\pm(F)}\, ,
\end{equation}
where $({}^\star G_{eff}^{-1})^{ij}$ is given in
Eq.(\ref{eq:zvezdaeffin}). Substituting these relations in the canonical Hamiltonian (\ref{eq:hamilton}), we obtain
\begin{equation}
T_{\pm}= \tilde T_{\pm} \, , \quad \mathcal{H}_{c}=\tilde
\mathcal{H}_{c}  \,  ,
\end{equation}
where we introduced an effective energy momentum tensor and Hamiltonian
\begin{equation}
\tilde T_{\pm}=\mp\frac{1}{4\kappa}\left[ ({}^\star
G_{eff}^{-1})^{ij} \;{}^\star \tilde j_{\pm i} \;{}^\star \tilde
j_{\pm j}+\frac{\alpha}{4} \;{}^\star \tilde j_{\pm (F)}
\;{}^\star \tilde j_{\pm (F)}\right]  \, , \quad \tilde
\mathcal{H}_{c}=\tilde T_- - \tilde T_+  \,  .
\end{equation}

The effective theory is defined in the phase space spanned by the
coordinates $q^i$ and momenta $p_i$ in
the new open string background $G_{ij}\to\; {}^\star
G^{eff}_{ij}$, $B_{ij}\to 0$, and $\Phi\to0$. The free field, which effective dynamics is described by ${}^\star f$ and $p$, decouples from the rest.

From the basic string variables algebra (\ref{eq:azs}), we calculate the corresponding effective
string one
\begin{equation}\label{eq:pzagrada}
\{q^{i}(\tau,\sigma),p_{j}(\tau,\overline{\sigma})\}=
{\delta^{i}}_{j}\delta_{s}(\sigma,\overline{\sigma}) \,  ,\quad
\{{}^\star f(\tau,\sigma),p(\tau,\overline{\sigma})\}=\delta_{s}(\sigma,\overline{\sigma})\,
,
\end{equation}
where
$\delta_{s}(\sigma,\overline\sigma)=\frac{1}{2}[\delta(\sigma-\overline\sigma)+\delta(\sigma+\overline\sigma)]$.

Separating the center of mass variables according to Appendix C, we obtain
\begin{equation}
\{ X^i(\sigma), X^j(\overline \sigma) \}={}^\star \Theta^{ij}\Delta(\sigma+\overline
\sigma)\, ,
\end{equation}
\begin{equation}
\{ X^i(\sigma), {}^\star \mathcal F(\overline \sigma)
\}=0\, ,\quad \{ {}^\star \mathcal F(\sigma),
{}^\star \mathcal F(\overline \sigma) \}=0\, ,
\end{equation}
where the function $\Delta(x)$ is given in Eq.(\ref{eq:DELTA}), and $X^i$ and ${}^\star \mathcal F$ are defined in (\ref{eq:cenmassX}) and (\ref{eq:cenmassF}), respectively.

The fields $x^i$ are noncommutative variables, while the field ${}^\star F$ is a commutative one. So, the Dp brane is described by $p+1$ noncommutative  and one commutative degree of freedom.

\section{Noncommutativity for singular ${}^\star G_{ij}$ ($\alpha a^2=1$)}

\setcounter{equation}{0}

In order to express the velocities in terms of the canonical momenta, the coefficients in front of the velocities must be different from zero. But the metric ${}^\star G_{ij}$ in front of $\dot x^i$ in (\ref{eq:kanimp}) is singular for $\alpha a^2=1$ [see Eqs.(\ref{eq:Lmetrika}) and (\ref{eq:detzvezda})]. Consequently, a primary constraint appears in the theory \cite{BN}. For $\alpha a^2=1$ the projector
\begin{equation}\label{eq:gmet}
g_{ij}=(P_T^0 G)_{ij}\, ,
\end{equation}
takes the role of the metric in the subspace defined by the regular part of ${}^\star G_{ij}$.

\subsection{Canonical Hamiltonian and gauge symmetry}

Combining the coordinates $x^i$ and their canonically conjugated momenta $\pi_i$ (\ref{eq:kanimp}) as
\begin{equation}
{}^\star j \equiv a^i \pi_i+2\kappa a^i B_{i j} {x^j}' = \kappa
(1-\alpha a^2)a_i \dot x^i\, ,
\end{equation}
we conclude that, for $\alpha a^2=1$, ${}^\star j$ does
not depend on velocities and consequently, it is a constraint of
the theory.

The canonical Hamiltonian $\mathcal{H}_c=\pi_i \dot
x^i+\pi\;{}^\star \dot F-\mathcal{L}$ in terms of currents has the form
\begin{equation}\label{eq:hamilton1}
\mathcal{H}_c= T_{-}-T_{+}\,  ,\quad  T_{\pm}=\mp\frac{1}{4\kappa}\left[ (g^{-1})^{ij}\;{}^\star j_{\pm i}\;{}^\star j_{\pm j}+\frac{\alpha}{4}\; {}^\star j_{\pm(F)} \;{}^\star  j_{\pm(F)}\right]\, ,
\end{equation}
where
\begin{equation}\label{eq:current2}
{}^\star j_{\pm i}=\pi_i+2\kappa (B_{ij}\pm\frac{1}{2}g_{ij})x'^j\, ,
\end{equation}
is obtained from (\ref{eq:zvezdastruje}) by imposing $\alpha a^2=1$, and
\begin{equation}
(g^{-1})^{ij}=(G^{-1}P_T^0)^{ij}\, ,
\end{equation}
is the metric inverse of (\ref{eq:gmet}) in the subspace defined by the regular part of ${}^\star G_{ij}$. 

The constraint ${}^\star j$ can be rewritten in terms of the current ${}^\star j_{\pm i}$ as
\begin{equation}
{}^\star j=a^i \;{}^\star j_{\pm i}\, .
\end{equation}
According to the Dirac theory for the constrained systems, we introduce the total Hamiltonian
\begin{equation}\label{eq:totalni}
H_T=\int d\sigma \mathcal{H}_T\, ,\quad
\mathcal{H}_T=\mathcal{H}_c+\lambda \;{}^\star j  \,  ,
\end{equation}
where  $\lambda$ is a Lagrange multiplier. From the current algebra (\ref{eq:algcur}) we have
\begin{equation}
\{{}^\star j_{\pm i}, {}^\star j\}=0  \,  ,\quad \{{}^\star j_{\pm(F)}, {}^\star
j\}=0\;\;\Longrightarrow  \{H_T, {}^\star j\}=0\, ,
\end{equation}
which means that ${}^\star j$ is a first class constraint. Consequently, there
is a gauge symmetry in the theory.

Using expression for the gauge transformation of
an arbitrary observable $X$, generated by symmetry generator $G$
\begin{equation}\label{eq:gtrans}
\delta_{\eta} X= \{X,G \}\, ,  \qquad   G \equiv \int d \sigma
\eta (\sigma) \;{}^\star j (\sigma)  \,  ,
\end{equation}
in this particular case we obtain
\begin{equation}\label{eq:prom1}
\delta_{\eta}x^i=a^i \eta \,   ,  \quad
\delta_{\eta}\;{}^\star F=0  \,  ,\qquad \delta_{\eta}\pi_i=2\kappa a^j B_{j i} \eta' \,  ,  \quad
\delta_{\eta}\pi=0  \,  .
\end{equation}

\subsection{Solution of constraints for particular gauge fixing}

From the gauge
transformations (\ref{eq:prom1}), it follows
\begin{equation}
\delta_\eta x_0\equiv \delta_\eta (a_i x^i)=a^2\eta\, ,
\end{equation}
and we conclude that $x_0=0$ is a good gauge condition. After
gauge fixing, we can treat ${}^\star j$ and $x_0$ as second class
constraints. Implementing the conditions $x_0=0$ and ${}^\star
j=0$, the current ${}^\star j_{\pm i}$ changes as
\begin{equation}
{}^\star j_{\pm i}\to j_{\pm i}=\pi_i+2\kappa \Pi_{\pm ij}x'^j\, ,\quad \left( \Pi_{\pm ij}=B_{ij}\pm \frac{1}{2}G_{ij}\right) 
\end{equation}
and the boundary conditions
(\ref{eq:gruslov}) take the form
\begin{equation}\label{eq:novigruslov}
\gamma_i^{(0)}=(\Pi_{+}G^{-1})_i{}^j j_{-\,j}+(\Pi_{-}G^{-1})_i{}^j j_{+\,j}\,
,\quad \gamma^{(0)}=\frac{1}{2}\left[ {}^\star
j_{-\,(F)}-\;{}^\star j_{+\,(F)}\right]\, .
\end{equation}
Like in the previous section, after Dirac consistency procedure, we obtain the $\sigma$-dependent form of the boundary conditions at $\sigma=0$
\begin{equation}
\Gamma_i(\sigma)=(\Pi_{+}G^{-1})_i{}^j j_{- j}(\sigma)+ (\Pi_{-}G^{-1})_i{}^j
j_{+ j}(-\sigma)\, ,\quad \Gamma(\sigma)=\frac{1}{2}\left[
{}^\star j_{-\,(F)}(\sigma)-\;{}^\star
j_{+\,(F)}(-\sigma)\right]\, .
\end{equation}
Similar expressions from the constraints at $\sigma=\pi$ are solved by periodicity of all variables.

Let us mark the complete set of the constraints with
$\chi_A=(\Gamma_i,\Gamma)$. The matrix form of the constraint
algebra is
\begin{equation}\label{eq:IIklasa}
\left\lbrace \chi_A(\sigma), \chi_B(\overline \sigma) \right\rbrace=-\kappa M_{AB}\delta'\, ,\qquad M_{AB}=\left (
\begin{array}{cc}
G^{eff}_{ij} & 0\\
0 & \frac{4}{\alpha}
\end{array}\right )\, ,
\end{equation}
where $G_{ij}^{eff}$ is defined after Eq.(\ref{eq:effm}).
Since we assume that $\det G^{eff}_{ij}\neq0$, we conclude that all constraints are of the second class.

In terms of the effective variables defined in Eq.(\ref{eq:mena1}), the constraint equations, $\Gamma_i(\sigma)=0$ and $\Gamma(\sigma)=0$, have the form
\begin{equation}\label{eq:prvagrupa}
\overline p_i=0\, ,\quad \overline q'^i=\frac{2}{\kappa}(G_{eff}^{-1}BG^{-1})^{ij}p_j\, ,
\end{equation}
\begin{equation}\label{eq:drugagrupa}
\overline p=0\, ,\quad {}^\star \overline f'=0\, .
\end{equation}
Because the first class constraints and gauge fixing behave like
second class constraints, from ${}^\star j=0$ and $x_0\equiv a_i x^i=0$, we get
additional equations
\begin{equation}\label{eq:fccs}
a^i p_i+2\kappa (aB)_i \overline q'^i=0\, ,\quad a^i \overline
p_i+2\kappa (aB)_i q'^i=0\, ,\quad q_0\equiv a_i q^i=0\, ,\quad
\overline q_0\equiv a_i \overline q^i=0\, .
\end{equation}
Combining the first equation in (\ref{eq:prvagrupa}) with the second one in (\ref{eq:fccs}) we have $q'_1\equiv (aB)_i q'^i=0$. The fourth equation in (\ref{eq:fccs}) and the second equation in (\ref{eq:prvagrupa}) give $p_1\equiv (\tilde a B)^i p_i=0$, while the second equation in (\ref{eq:prvagrupa}) and the first one in (\ref{eq:fccs}) produce $p_0\equiv \tilde a^i p_i=0$. From here we conclude that the string phase space is spanned by the following coordinates and momenta
\begin{equation}\label{eq:var2}
(q_T)^i=({}^\star P_T)^i{}_j q^j\equiv Q^i\, ,\quad (\pi_T)_i=({}^\star P_T)_i{}^j p_j\equiv P_i\, ,
\end{equation}
where the projector ${}^\star P_T$ is defined in (\ref{eq:starpt}) for $\alpha \tilde a^2\neq1$ and $\tilde a^2\neq0$. Decomposing $\overline q^i$ into a directions along $(aB)_i$ and $a_i$ and the orthogonal ones
\begin{equation}
\overline q'^i=\overline q_0'^i+\overline q_1'^i+(\overline q_T)'^i\, , 
\end{equation}
from the second equation (\ref{eq:prvagrupa}) we obtain
\begin{equation}
(\overline q_T)'^i=-2\;{}^\star \Theta^{ij}P_j\, ,\quad \overline q_1'^i=\frac{2}{\kappa}(G_{eff}^{-1}\;{}^\star P_1 BG^{-1})^{ij}P_j\, ,
\end{equation}
where the tensor
\begin{equation}\label{eq:tetka2}
{}^\star \Theta^{ij}=-\frac{1}{\kappa}( G_{eff}^{-1}\;{}^\star P_T B G^{-1}\;{}^\star P_T)^{ij}\, ,
\end{equation}
is antisymmetric.

Choosing the integration constants $q_1^i(\sigma=0)=0$, $\overline q^i(\sigma=0)=0$, and ${}^\star \overline f(\sigma=0)=0$, the final solution of the Eqs.(\ref{eq:prvagrupa}), (\ref{eq:drugagrupa}), and (\ref{eq:fccs}) takes the form
\begin{equation}\label{eq:drugoresenje}
x^i_{D_p}(\sigma)=Q^i(\sigma)-2\;{}^\star \Theta^{ij}\int^{\sigma}_0 d\sigma_1 P_j(\sigma_1)\, ,\quad  \pi_i^{D_p}=P_i\, ,
\end{equation}
\begin{equation}\label{eq:x1pi1}
x_0=0\, , \quad \pi_0=0\, ,\quad x_1(\sigma)=\frac{2}{\kappa}(\tilde aB^2G^{-1})^i\int^\sigma_0 d\sigma_1 P_i(\sigma_1)\, , \quad \pi_1=0\, ,
\end{equation}
\begin{equation}\label{eq:resenjeF}
{}^\star F={}^\star f\, ,\quad \pi=p\, .
\end{equation}
Similar as in (\ref{eq:var2}), we introduced the notation 
\begin{equation}\label{eq:dpvar}
x^i_{D_p}=({}^\star P_T)^i{}_j x^j\, ,\quad \pi_i^{D_p}=({}^\star P_T)_i{}^j \pi_j\, , 
\end{equation}
while $x_0$, $x_1$, $\pi_0$, and $\pi_1$ are defined in Eq.(\ref{eq:notacija}). The solution for $x_1$  satisfies Dirichlet boundary conditions at $\sigma=\pi$, $x_1(\sigma=\pi)=0$, as a consequence of the periodicity of the momenta $P_i$.

\subsection{Effective theory and noncommutativity}

In terms of the effective currents
\begin{equation}
{}^\star \tilde j_{\pm i}=P_i\pm \kappa ({}^\star P_T G_{eff})_{ij}Q'^j\, ,\quad {}^\star \tilde j_{\pm(F)}=p\pm \frac{4\kappa}{\alpha}\;{}^\star f'\, ,
\end{equation}
the currents ${}^\star j_{\pm i}$ and ${}^\star j_{\pm(F)}$ given in (\ref{eq:current2}) and (\ref{eq:zvezdastruje}) can be expressed as
\begin{equation}
{}^\star j_{\pm i}=\pm 2 (\Pi_{\pm}G_{eff}^{-1})_i{}^j {}^\star \tilde j_{\pm j}\, , \quad {}^\star j_{\pm(F)}={}^\star \tilde j_{\pm(F)}\, .
\end{equation}
Substituting these relations in (\ref{eq:hamilton1}), we obtain the effective Hamiltonian
\begin{equation}
\tilde \mathcal H_c=\tilde T_{-}-\tilde T_{+}\, ,\quad \tilde
T_{\pm}=\mp\frac{1}{4\kappa}\left[ (G_{eff}^{-1}\;{}^\star P_T)^{ij}\;{}^\star \tilde j_{\pm i}\;{}^\star \tilde j_{\pm j}+\frac{\alpha}{4}\;{}^\star \tilde j_{\pm (F)}\;{}^\star \tilde j_{\pm (F)}\right]\, .
\end{equation}

The expressions for the effective current ${}^\star \tilde j_{\pm i}$ and the energy-momentum tensor $\tilde T_{\pm}$ show that the effective metric and its inverse are of the form
\begin{equation}
g^{eff}_{ij} = ({}^\star P_T G^{eff})_{ij} \,  ,  \qquad   g_{eff}^{ij} = (G_{eff}^{-1}\;{}^\star P_T)^{ij} \, .
\end{equation}
Therefore, the string propagates in the subspace defined by the projector ${}^\star P_T$ in the background
\begin{equation}
G_{ij}\to g^{eff}_{ij}\, ,\quad B_{ij}\to
0\, ,\quad \Phi\to 0\, . 
\end{equation} 
The effective dynamics of the string is described by the effective variables: coordinates $Q^i$ and momenta $P_i$, which satisfy the
algebra
\begin{equation}
\left\lbrace Q^i(\sigma), P_j(\overline\sigma)\right\rbrace=({}^\star P_T)^i{}_j \delta(\sigma-\overline\sigma)\, .
\end{equation}
The conformal part of the effective world sheet metric ${}^\star f$ and its momentum $p$ are canonical variables for the scalar degree of freedom which decouples from the rest.

Using the solutions (\ref{eq:drugoresenje}) and (\ref{eq:resenjeF}), and introducing the center of mass variables according to the Appendix C, the noncommutativity relations take the form
\begin{equation}
\{ X_{D_p}^i(\sigma), X_{D_p}^j(\overline \sigma)\}={}^\star \Theta^{ij}\Delta(\sigma+\overline \sigma)\, ,
\end{equation}
\begin{equation}
\left\lbrace X_{D_p}^i(\tau,\sigma),{}^\star \mathcal F(\tau,\overline\sigma) \right\rbrace=0\, ,\quad \left\lbrace {}^\star \mathcal F(\tau,\sigma),{}^\star \mathcal F(\tau,\overline\sigma) \right\rbrace=0  \,  ,
\end{equation}
where the tensor ${}^\star \Theta^{ij}$ and the function $\Delta(x)$ are defined in (\ref{eq:DELTA}) and (\ref{eq:tetka2}).

The solutions for $x_0$ and $x_1$  satisfy the Dirichlet boundary conditions and decrease the number of the Dp-brane dimensions from $p+2$ to $p$.
There is one commutative
variable, the conformal part of the intrinsic metric ${}^\star F$, and
$p-1$ noncommutative ones $x^i_{D_p}$.

\section{Noncommutativity for singular ${}^\star G^{eff}_{ij}$ ($\alpha \tilde a^2=1$)}
\setcounter{equation}{0}

For $\tilde A\equiv 1-\alpha \tilde a^2=0$ and $A\equiv 1-\alpha a^2\neq0$ the complete canonical analysis as well as the consistency
procedure for the constraints, performed in Sec.III, can be
repeated here. The difference appears in the separation of the first from the second class constraints as a consequence of the singulatity of matrix $M_{AB}$ (\ref{eq:detM}).

\subsection{From the second to the first class constraints}

Using the expression for effective metric (\ref{eq:zvezdaeff}), we obtain
\begin{equation}
{}^\star G^{eff}_{ij}\tilde a^j=\tilde A a_i\, ,\quad {}^\star
G^{eff}_{ij} (\tilde a B)^j=\frac{\tilde A}{A} (aB)_i\, ,
\end{equation}
so that, for $\tilde A=0$ and $A\neq0$,
$\tilde a^i$ and $(\tilde a B)^i$ are singular vectors of the metric ${}^\star G_{ij}^{eff}$. According to Eq.(\ref{eq:dklasa}) we expect that two constraints originating from the boundary conditions turn into the first class.

In order to investigate the theory with constraints, we introduce the total Hamiltonian
\begin{eqnarray}
H_T=\int d\sigma \mathcal{H}_T \, ,\quad
\mathcal{H}_T=\mathcal{H}_c+\lambda^i(\sigma)\Gamma_i(\sigma)+\lambda(\sigma) \Gamma(\sigma)\, ,
\end{eqnarray}
where $\mathcal H_c$ is defined in (\ref{eq:hamilton}), $\Gamma_i$ and $\Gamma$ are defined in (\ref{eq:velikog}), and $\lambda^i$ and $\lambda$ are Lagrange multipliers. We decompose $\lambda ^i$ using the projectors $({}^\star \hat P_0)_i{}^j$, $({}^\star \hat P_1)_i{}^j$,  and $({}^\star \hat P_T)_i{}^j$, defined in the appendix A
\begin{equation}\label{eq:razlaganje}
\lambda^i=(\lambda_T)^i+2\Lambda_1(\tilde
aB)^i+\Lambda_2\tilde a^i\, ,
\end{equation}
where $(\lambda_T)^i=({}^\star \hat P_T)^i{}_j \lambda^j$,
$\Lambda_1=-\frac{2\alpha}{1-\alpha a^2}(a B\lambda)$, and
$\Lambda_2=\alpha (a \lambda)$. The consistency conditions,
$\{H_T,\Gamma_i(\sigma)\}\approx0$ and $\{H_T,\Gamma(\sigma)\}\approx0$,
enable us to calculate the coefficients
\begin{equation}\label{eq:okof}
\lambda'=-\frac{\alpha}{4\kappa}\Gamma'\, ,\quad
(\lambda_T')^i=-\frac{1}{\kappa}(G^{-1}_{eff}\;{}^\star \hat
P_T)^{ij}\Gamma'_j\, ,
\end{equation}
while the coefficients
$\Lambda_1$ and $\Lambda_2$ remain undetermined.

The total Hamiltonian takes the form
\begin{equation}\label{eq:popham}
H_T=H_c+\int_0^\pi d\sigma \left[ (\lambda_T)^i(
\Gamma_T)_i+\lambda\Gamma+\Lambda_1 \Gamma_1+\Lambda_2
\Gamma_2\right] =H'+\int_0^\pi d\sigma (\Lambda_1
\Gamma_1+\Lambda_2 \Gamma_2)\, ,
\end{equation}
where
\begin{equation}\label{eq:prvi}
\Gamma_1=2(\tilde a BG^{-1})^i \Gamma_i\, ,\quad \Gamma_2=\tilde
a^i \Gamma_i\, .
\end{equation}
Since the constraints $\Gamma_1$ and $\Gamma_2$ are multiplied by
the arbitrary coefficients $\Lambda_1$ and $\Lambda_2$, they are
of the first class. On the other hand, $(\Gamma_T)_i$ and
$\Gamma$, multiplied by the determined multipliers, are of the
second class.

By direct calculation, from (\ref{eq:dklasa}), we have
\begin{equation}\label{eq:nule1}
\left\lbrace \Gamma_1,\Gamma_i\right\rbrace =0\, ,\quad \left\lbrace
\Gamma_1,\Gamma\right\rbrace =0\, , \qquad \left\lbrace
\Gamma_2,\Gamma_i\right\rbrace =0\, ,\quad \left\lbrace
\Gamma_2,\Gamma\right\rbrace =0\, ,
\end{equation}
which is a confirmation that $\Gamma_1$ and $\Gamma_2$
are of the first class.

Calculating the algebra of the constraints $\chi_A=\left\lbrace (\Gamma_T)_i, \Gamma \right\rbrace $ we obtain
\begin{equation}
\left\lbrace \chi_A,\chi_B\right\rbrace =-\kappa M_{AB} \delta'\, ,\qquad M_{AB}=\left(
\begin{array}{cc}
({}^\star \hat P_T G_{eff})_{ij} & 0 \\ 0 & \frac{4}{\alpha}
\end{array}  \right)\, .
\end{equation}
Because the projector $({}^\star \hat P_T)_i{}^j$ is orthogonal to
the vectors $a_i$ and $(aB)_i$, we conclude that the rank of
$M_{AB}$ is not greater than $p$. Assuming that the rest of the
matrix $M_{AB}$ is regular, its rank as well as the
number of the second class constraints are equal to $p$.

\subsection{Gauge symmetry and solution of constraints}

The gauge transformations have the form of the Eq.(\ref{eq:gtrans}), with the generator
\begin{equation}\label{eq:localsym}
G=\int_0^\pi d\sigma (\eta_1\Gamma_1+\eta_2\Gamma_2)\, ,
\end{equation}
where $\eta_1$ and $\eta_2$ are the parameters of the local transformations. The constraints
\begin{equation}\label{eq:fc}
\Gamma_1=\tilde a^i p_i+2(\tilde aBG^{-1})^i\overline p_i\,
,\quad \Gamma_2=\tilde a^i \overline p_i+2(\tilde
aBG^{-1})^i p_i\, ,
\end{equation}
generate the gauge transformations of the effective variables
\begin{equation}\label{eq:g1}
\delta q^i=\tilde a^i(\eta_1)_s+2(\tilde aBG^{-1})^i(\eta_2)_s\,
,\quad \delta \;{}^\star f=0\, ,
\end{equation}
\begin{equation}\label{eq:g2}
\delta \overline q^i=\tilde a^i(\eta_2)_a+2(\tilde
aBG^{-1})^i(\eta_1)_a\, ,\quad \delta \;{}^\star \overline f=0\, ,
\end{equation}
where the indices $"s"$ and $"a"$ denote $\sigma$ symmetric and antisymmetric
parts of the parameters $\eta_1$ and $\eta_2$. The particular gauge transformations
\begin{equation}
\delta q_0=\tilde a^2 \eta_{1s}\, ,\quad \delta \overline q_0=\tilde a^2 \eta_{2a}\, ,\quad \delta q_1=\frac{1}{2\alpha}(\alpha a^2-1)\eta_{2s}\, ,\quad \delta \overline q_1=\frac{1}{2\alpha}(\alpha a^2-1)\eta_{1a}\, ,
\end{equation}
enable us to choose good gauge fixing
\begin{equation}\label{eq:gejdz}
q_0=0\, ,\quad \overline q_0=0\, ,
\quad q_1=0\, ,\quad \overline q_1=0\, .
\end{equation}

Now, the first class constraints and gauge conditions behave like second class constraints. So, the full set of expressions, $\Gamma_i$ and $\Gamma$  (\ref{eq:veza1}), vanishes as second class constraints.

Choosing the integration constants $\overline q^i(\sigma=0)=0$ and
${}^\star \overline f(\sigma=0)=0$,  from $\Gamma_i=0$, $\Gamma=0$, and
gauge conditions (\ref{eq:gejdz}), we get the solution
\begin{equation}\label{eq:resenje12}
x^i_{D_p}(\sigma)=\hat Q^i(\sigma)-2\;{}^\star
\Theta^{ij}\int^\sigma_0 d\sigma_1 \hat P_j(\sigma_1)\, ,\quad
\pi_i^{D_p}=\hat P_i \,  ,
\end{equation}
\begin{equation}\label{eq:Fresenje}
x_0=0\, ,\quad \pi_0=0\, ,\quad x_1=0\, ,\quad \pi_1=0\, ,\quad
{}^\star F={}^\star f\, ,\quad \pi=p\, .
\end{equation}
For $\hat Q^i$, $\hat P_i$, $x^i_{D_p}$, and $\pi_i^{D_p}$ we used the similar notation as in (\ref{eq:var2}) and (\ref{eq:dpvar})
\begin{equation}
(q_T)^i=({}^\star \hat P_T)^i{}_j q^j\equiv \hat Q^i\, ,\quad (p_T)_i=({}^\star \hat P_T)_i{}^j p_j\equiv \hat P_i\, ,\quad x^i_{D_p}=({}^\star \hat P_T)^i{}_j x^j\, ,\quad \pi_i^{D_p}=({}^\star \hat P_T)_i{}^j \pi_j\, ,
\end{equation}
but now using the projector $({}^\star \hat P_T)_i{}^j$ instead of $({}^\star P_T)_i{}^j$. The vector components $x_0\, ,x_1\, ,\pi_0$, and $\pi_1$ are
introduced in Eq.(\ref{eq:notacija}), and the tensor ${}^\star \Theta^{ij}$
\begin{equation}\label{eq:teta5}
{}^\star \Theta^{ij}=-\frac{1}{\kappa}(G_{eff}^{-1}\;{}^\star \hat P_T B G^{-1}\;{}^\star \hat P_T)^{ij}\, ,
\end{equation}
is manifestly antisymmetric.

\subsection{Effective theory}

Let us introduce the effective currents
\begin{equation}
{}^\star \tilde j_{\pm i}=\hat P_i \pm \kappa ({}^\star \hat P_T
G_{eff})_{ij}\hat Q'^j\, ,\quad {}^\star \tilde
j_{\pm(F)}=p\pm\frac{4\kappa}{\alpha}\;{}^\star f'\, ,
\end{equation}
and correlate them with the currents defined in Eq.(\ref{eq:zvezdastruje})
\begin{equation}\label{eq:efekat2}
{}^\star j_{\pm i}=\pm 2(\Pi_{\pm} G_{eff}^{-1})_i{}^j \;{}^\star
\tilde j_{\pm j}-4(\Pi_{\pm}G_{eff}^{-1}\;{}^\star \hat P_1
B)_i{}^j \hat P_j\, ,\quad {}^\star j_{\pm(F)}={}^\star \tilde
j_{\pm(F)}\, .
\end{equation}

Substituting these relations in the expression for
energy-momentum tensor (\ref{eq:hamilton}) we obtain
\begin{equation}
T_{\pm}=\tilde T_{\pm}\, ,\quad  \mathcal H_c==\tilde T_{-}-\tilde T_{+}\equiv \tilde \mathcal{H}_c\, ,
\end{equation}
where
\begin{equation}
\tilde T_{\pm}=\mp \frac{1}{4\kappa}\left[ (G_{eff}^{-1}\;{}^\star
\hat P_T)^{ij}\;{}^\star\tilde j_{\pm i}\;{}^\star\tilde j_{\pm
j}+\frac{\alpha}{4}\;{}^\star \tilde j_{\pm(F)} {}^\star \tilde
j_{\pm(F)}\right] \, .
\end{equation}

The effective theory lives in the
background $G_{ij} \to g^{eff}_{ij}=({}^\star \hat P_T
G_{eff})_{ij}, \, B_{ij} \to 0$, and $\Phi\to0$. The string dynamics is described
by the effective variables $\hat Q^i$, $\hat P_j$, ${}^\star f$, and
$p$.

\subsection{Noncommutativity}

From the algebra (\ref{eq:pzagrada}), we
obtain the algebra of the effective variables
\begin{equation}
\left\lbrace \hat Q^i(\sigma), \hat P_j(\overline \sigma)
\right\rbrace= ({}^\star \hat P_T)^i_{\;j}
\delta_s(\sigma,\overline \sigma)\, ,
\end{equation}
where $\delta_s(\sigma,\overline \sigma)$ is defined after Eq.(\ref{eq:pzagrada}).

As in the two previous cases, ${}^\star F$ is decoupled and takes
the role of the commutative variable. Introducing the center of mass
variables according to Appendix C, with the help of
the Eqs.(\ref{eq:resenje12}), we have
\begin{equation}
\{X^{i}_{D_p}(\tau,\sigma),X^{j}_{D_p}(\tau,\overline{\sigma})\}={}^\star
\Theta^{ij}\Delta(\sigma+\overline{\sigma})\, ,
\end{equation}
where the antisymmetric tensor ${}^\star \Theta^{ij}$ is given in
Eq.(\ref{eq:teta5}).

It follows from (\ref{eq:resenje12}) and (\ref{eq:Fresenje}) that
$x_0$ and $x_1$ are fixed and, consequently, satisfy Dirichlet
boundary conditions and decrease the number of Dp-brane
dimensions. All other $p-1$ Dp-brane coordinates are
noncommutative.

\section{Concluding remarks}
\setcounter{equation}{0}

In this article we used the possibility to establish the conformal invariance adding the Liouville term to the action, instead to use the third space-time equation of motion, $\beta^\Phi=0$. We showed that this change preserves main results of the previous paper \cite{PR}: (1) existence of the local gauge symmetries, which decrease the number of the Dp-brane dimensions; (2) the number of the commutative and noncommutative variables.

In fact, the Liouville action cancels the remaining constant anomaly, $\beta^\Phi=c$, after imposing the first two space-time equations of motion, $\beta^G_{\mu\nu}=0=\beta^B_{\mu\nu}$. It also makes the conformal part of the world sheet metric, $F$, dynamical variable. The theory becomes bilinear in $F$, with the quadratic Liouville term and linear term with the dilaton field. It is easy to change the variables, $F\to{}^\star F=F+\frac{\alpha}{2}a_i x^i$, so that term linear in $F$ disappears. As a consequence, the quadratic term in $x^i$ appears which change the metric tensor, $G_{ij}\to{}^\star G_{ij}=G_{ij}-\alpha a_i a_j$. For particular values of the square of the vector $a_i$, with respect to the closed string metric, $a^2=\frac{1}{\alpha}$, and to the effective one, $\tilde a^2=\frac{1}{\alpha}$, the corresponding star metrics become singular [see Eqs. (\ref{eq:detzvezda}) and (\ref{eq:detzvezdaeff})]

We analyzed three cases: (1) $\alpha a^2\neq1\,
,\alpha \tilde a^2\neq1$, (2) $\alpha a^2=1\,
,\alpha \tilde a^2\neq1$, and (3) $\alpha \tilde
a^2=1\, ,\alpha a^2\neq1$. In all cases the
field ${}^\star F$ decouples, so it is a commutative variable. The
rest part of the action formally has the same form as in the dilaton
free case, where the regular metric $G_{ij}$ is substituted by the metric ${}^\star G_{ij}$, which can be singular for some choices of the background fields. The case (1) corresponds to such values of parameters that the
star metric ${}^\star G_{ij}$ is regular. So, everything bahaves like in the dilaton free case. In particular, all Dp-brane
coordinates $x^i$ are noncommutative.

The singularities of the star metrics have different influences to the canonical constraints. In the case (2), ${}^\star G_{ij}$ is coefficient in front of the velocity $\dot x^i$, so its singularity produces standard canonical constraint. In the case (3), the algebra of the constraints originating from boundary conditions, closed on ${}^\star G_{ij}^{eff}$. So, the singularity of ${}^\star G_{ij}^{eff}$ changes the character of the constraints, turning some of them from the second to the first class. According to Appendix B, ${}^\star G_{ij}$ has one singular direction $a^i$, while ${}^\star G_{ij}^{eff}$ has two singular directions, $\tilde a^i$ and $(\tilde aB)^i$. Therefore, in the case (2) there is one and in the case (3) there are two first class constraints.  

In both cases the first class contraints are the symmetry
generators. After the gauge fixing, gauge conditions and the first class
constraints can be considered as second class constraints. Solving
all second class constraints (both the original ones and the first
class constraints with gauge conditions), we obtain the string coordinates in terms of the effective ones.

The solutions (\ref{eq:solution1}), (\ref{eq:drugoresenje}), and
(\ref{eq:resenje12}) have the same general form
\begin{equation}
x^i_{D_p}(\sigma)=Q^i(\sigma)-2\;{}^\star \Theta^{ij}\int^\sigma_0
d\sigma_1 P_{j}(\sigma_1)\, .
\end{equation}
The string coordinates $x^i_{Dp}=({}^\star P_{D_p})^i{}_j x^j$ are expressed in terms of the
effective canonical variables
\begin{equation}
Q^i=({}^\star P_{D_p})^i{}_j q^j\, ,\quad P_{i}=({}^\star
P_{D_p})_i{}^j p_j\, ,
\end{equation}
satisfying the algebra
\begin{equation}
\left\lbrace
Q^i(\sigma),P_j(\overline\sigma)\right\rbrace=({}^\star
P_{D_p})^i{}_j\delta_s(\sigma,\overline\sigma)\, .
\end{equation}

In
the second and third case, the string coordinates $x_0\equiv(n_0)_i x^i=a_i x^i$ and
$x_1\equiv(n_1)_i x^i=(aB)_i x^i$ satisfy Dirichlet boundary conditions and decrease the number of the Dp-brane dimensions. It is known that boundary conditions are usually imposed on space-like variables. Because the coordinates $x_0$ and $x_1$ satisfy Dirichlet boundary conditions, it is important to clarify the nature of the vectors $(n_0)_i$ and $(n_1)_i$. Let us first introduce explicite dependence on the string slope parameter $\alpha'=\frac{1}{2\pi\kappa}$, by simple redefinition of dilaton field, $\Phi\to\alpha'\Phi$. Then the singularities of metrics ${}^\star G_{ij}$ and ${}^\star G^{eff}_{ij}$ occur at $a^2=\frac{1}{\alpha \alpha'^2}$ and $\tilde a^2=\frac{1}{\alpha \alpha'^2}$, respectively, and from (\ref{eq:betaFis}) and (\ref{eq:alfa}) we obtain 
\begin{equation}
\frac{1}{\alpha \alpha'^2}=\frac{\beta^\Phi}{4}=\frac{D-26}{24\alpha'}+a^2\, .
\end{equation}
From the first singularity condition the $a^2$ dependence disappears and we obtain that string must be critical, $D=26$. Because there are no conditions on $(n_0)_i$ and $(n_1)_i$, we can choose them to be space-like variables, $n_0^2=a^2>0$ and $n_1^2=-(a B^2 a)>0$. From the second singularity condition, with the help of the relation $\tilde a^2-a^2=4\tilde a B^2 a=4 aB^2a+16\tilde a B^4 a$, we obtain the conditions for the vectors $(n_0)_i$ and $(n_1)_i$ to be space-like
\begin{equation}
n_0^2=\tilde a^2-\frac{D-26}{24\alpha'}>0\, ,\qquad n_1^2=4\tilde a B^4 a-\frac{D-26}{96\alpha'}>0\, .
\end{equation}
For $D\leqslant 26$, in order to satisfy these conditions, it is enough to choose $\tilde a^2\geqslant0$ and $\tilde a B^4 a\geqslant0$.

The string components, $x_{D_p}^i$ are noncommutative degrees of freedom, because they depend on the effective coordinates and momenta. The
noncommutativity relation between the Dp-brane coordinates has the
same form in all three cases
\begin{equation}\label{eq:PB}
\{ X^i_{D_p}(\tau,\sigma),X^j_{D_p}(\tau,\overline
\sigma)\}={}^\star \Theta^{ij}\Delta(\sigma+\overline\sigma)\, ,
\quad \Big[
X^i_{D_p}(\sigma)=x^i_{D_p}(\sigma)-(x^i_{D_p})_{cm}\Big]\, .
\end{equation}
The interior of the string is
commutative and noncommutativity occurs on the string endpoints.
The noncommutativity parameter ${}^\star \Theta^{ij}$ in the first
case is given in (\ref{eq:tetka1}), while in the other two cases
it can be expressed in terms of the projectors ${}^\star P_{D_p}$
\begin{equation}
{}^\star \Theta^{ij}=-\frac{1}{\kappa}(G_{eff}^{-1} \;{}^\star
P_{D_p} BG^{-1}\;{}^\star P_{D_p})^{ij}\, .
\end{equation}
All important results of this analysis are presented in Table 1 \vspace{0.5cm}

\begin{table}[h]
\begin{tabular}{|c|c|c|c|c|c|c|}\hline
\textbf{Case} & $D_{Dp}$ &
$({}^\star P_{D_p})_i{}^j$ & $V_{Dbc}$ & $x_{D_p}^i$ & $g_{ij}^{eff}$\\
\hline\hline $\alpha \tilde a^2\neq1$,
$\alpha a^2\neq1$ & p+2 & $\delta_i{}^j$ & \dots & $x^i$ & ${}^\star G_{ij}^{eff}$\\ \hline
$\alpha a^2=1\, ,\alpha \tilde a^2\neq1$ & p &
$({}^\star P_T)_i{}^j$ &
$x_0\, ,x_1$ & $({}^\star P_T x)^i$ & $({}^\star P_T G_{eff})_{ij}$ \\
\hline $\alpha \tilde a^2=1\, ,\alpha a^2\neq1$ & p & $({}^\star \hat P_T)_i{}^j$ & $x_0\, ,x_1$ & $({}^\star \hat P_T x)^i$ & $({}^\star \hat P_T G_{eff})_{ij}$\\
\hline
\end{tabular}
\caption{Dp-brane features dependence on background fields}
\end{table}
\noindent where $D_{Dp}$ is the number of the Dp-brane dimensions, the
symbol $V_{Dbc}$  is related to variables with Dirichlet
boundary condition, and the
effective metrics are denoted by $g_{ij}^{eff}$. All projectors
are defined in Appendix A.

Let us stress that the solution of the boundary conditions, the number of the Dp-brane
dimensions, the number of the commutative and noncommutative
coordinates as well as the form of the noncommutativity parameter, in the approach with the Liouville
action are the same as in the approach presented in Ref.\cite{PR}. There are two formal differences. When we deal with the Liouville action, the
gauge symmetries appear for $\alpha a^2 =1$ and $\alpha \tilde
a^2 =1$ when star metrics, ${}^\star G_{ij}$ and ${}^\star G_{ij}^{eff}$, are singular instead for $a^2=0$ and $\tilde a^2 =0$ in the absence of the Liouville term. Also some commutative and noncommutative variables switched the roles, $x_0\to\;{}^\star F$
and $F\to x_0$. 

The inclusion of the Lioville term produces few advantages. First, there are only two space-time equations of motion (originated from $\beta_{\mu\nu}^G=0$ and $\beta_{\mu\nu}^B=0$) instead of three ones without Liouville. Second, presence of $F$ does not break the closed string conformal invariance. Consequently, there is no possibility that $F$-dependent open string boundary conditions break this invariance and there are no needs for additional restrictions on background fields, as in the absence of Liouville term. Finally, the complete solution including noncommutative parameter and effective variables depend on additional parameter, the central charge $c$.

It is interesting to mention that the effect of boundary conditions reduces the dimension of Dp brane by 2, from $p$ to $p-2$, as well as double T-duality. In fact any T-duality relates Dp brane wrapped around compact direction with radius $R$ to the D(p-1)-brane with dual compact radius $\tilde R$. So, two T-dualities along $x_0=a_i x^i$ and $x_1=(aB)_i x^i$ with compactification radii $R_0$ and $R_1$, could transform Dp brane to D(p-2)-brane with compactified radii $\tilde R_0$ and $\tilde R_1$. Possible deeper understanding of our result in terms of T-dualities is under investigation.

\appendix

\section{Projectors}
\setcounter{equation}{0}

In this appendix we introduce projector operators in order to
separate noncommutative and nonphysical variables on the Dp brane
as well as to express the noncommutativity parameter.

The projectors on the direction $n_i$ and on the subspace orthogonal to vector $n_i$ are
\begin{equation}
(\Pi)_i{}^j=\frac{n_i n^j}{n^2}\, ,\quad (\Pi_T)_i{}^j=\delta_i{}^j-(\Pi)_i{}^j\, ,
\end{equation}
where $n^i=g^{ij}n_j$ and $n^2=n^i n_i$ . The transposed operator is
\begin{equation}\label{eq:trans}
\Pi^i{}_j=g^{ik}\Pi_k{}^l g_{lj}\, .
\end{equation}

\subsection{Case $n_i=a_i$ and $g_{ij}=G_{ij}$}

For $n_i\to (n_0)_i=a_i$ and $g_{ij}\to G_{ij}$ we obtain
\begin{equation}\label{eq:ptpl}
(\Pi)_i{}^j\to(P_0)_i{}^j=\frac{a_i a^j}{a^2}\, ,\quad (\Pi_T)_i{}^j\to(P_T^0)_i{}^j=\delta_i{}^j-(P_0)_i{}^j\, .
\end{equation}

\subsection{Case $(n_0)_i=a_i$ and $(n_1)_i=(aB)_i$ and $g_{ij}=G^{eff}_{ij}$}

Let us construct the projector orthogonal to the vectors $(n_0)_i=a_i$ and $(n_1)_i=(aB)_i$ with respect to the effective metric $G_{ij}^{eff}$. These two vectors are mutually orthogonal and it is enough to use the projectors on the direction $a_i$
\begin{equation}\label{eq:pitenula}
(\Pi_0)_i{}^j=\frac{a_i \tilde a^j}{\tilde a^2}\, ,
\end{equation}
and on the direction $(aB)_i$
\begin{equation}\label{eq:pipit1}
(\Pi_1)_i{}^j=\frac{4}{\tilde a^2-a^2}(Ba)_i(\tilde aB)^j\, ,
\end{equation}
to construct the projector orthogonal on them
\begin{equation}\label{eq:pite}
(\Pi_T)_i{}^j=\delta_i{}^j-(\Pi_0)_i{}^j-(\Pi_1)_i{}^j\, .
\end{equation}

In the case when $\alpha a^2=1$ we have
\begin{equation}\label{eq:starpt1}
({}^\star P_1)_i{}^j=(\Pi_1)_i{}^j\Big|_{\alpha a^2=1}=\frac{4\alpha}{\alpha \tilde a^2-1}(Ba)_i(\tilde aB)^j\, , 
\end{equation}
\begin{equation}\label{eq:starpt}
({}^\star P_T)_i{}^j=(\Pi_T)_i{}^j\Big|_{\alpha a^2=1}=\delta_i{}^j-({}^\star P_0)_i{}^j-({}^\star P_1)_i{}^j\, ,
\end{equation}
where by definition we put
\begin{equation}
({}^\star P_0)_i{}^j=(\Pi_0)_i{}^j=\frac{a_i \tilde a^j}{\tilde a^2}\, .
\end{equation}
Similarly for $\alpha \tilde a^2=1$ we get
\begin{equation}
({}^\star \hat P_0)_i{}^j=(\Pi_0)_i{}^j\Big|_{\alpha \tilde a^2=1}=\alpha a_i \tilde a^j\, ,\quad ({}^\star \hat P_1)_i{}^j=(\Pi_1)_i{}^j\Big|_{\alpha \tilde a^2=1}=\frac{4\alpha}{1-\alpha a^2}(Ba)_i(\tilde aB)^j\, ,
\end{equation}
\begin{equation}\label{eq:starptkapa}
\quad ({}^\star \hat P_T)_i{}^j=(\Pi_T)_i{}^j\Big|_{\alpha \tilde a^2=1}=\delta_i{}^j-({}^\star \hat P_0)_i{}^j-({}^\star \hat P_1)_i{}^j\, .
\end{equation}

An arbitrary contravariant vector $x^i$ decomposes as
\begin{equation}
x^i=(x_0)^i+(x_1)^i+(x_T)^i\, ,\quad (x_0)^i=(\Pi_0)^i{}_j x^j\, ,\quad (x_1)^i=(\Pi_1)^i{}_j x^j\, , \quad (x_T)^i=(\Pi_T)^i{}_j x^j\, ,
\end{equation}
as well as an arbitrary covariant vector $\pi_i$
\begin{equation}
\pi_i=(\pi_0)_i+(\pi_1)_i+(\pi_T)_i\, ,\quad (\pi_0)_i=(\Pi_0)_i{}^j \pi_j\, ,\quad (\pi_1)_i=(\Pi_1)_i{}^j \pi_j\, , \quad (\pi_T)_i=(\Pi_T)_i{}^j \pi_j\, .
\end{equation}

It is useful to introduce the following notation for the projections of vectors $x^i$ and $\pi_i$
\begin{eqnarray}
x_0&=&(n_0)_i x^i=a_i x^i\, ,\quad x_1=(n_1)_i x^i=(aB)_i x^i\,
,\nonumber \\ \pi_0&=&\tilde n_0^i \pi_i=\tilde a^i \pi_i\,
,\qquad \pi_1=\tilde n_1^i \pi_i=(\tilde aB)^i
\pi_i\label{eq:notacija}\, .
\end{eqnarray}

\section{The star metrics ${}^\star G_{ij}$ and ${}^\star
G^{eff}_{ij}$} \setcounter{equation}{0}

Here we are going to introduce the expressions for the redifined metrics in the presence of the Liouville action,
${}^\star G_{ij}$ and ${}^\star
G^{eff}_{ij}$. The metric ${}^\star G_{ij}$ is defined as
\begin{equation}\label{eq:Lmetrika}
{}^\star G_{ij}=G_{ij}-\alpha a_i a_j=(P_T^0+A\,P_0)_i{}^k
G_{kj}\, ,\quad A=1-\alpha a^2 \, ,
\end{equation}
while, for $A\neq0$, its inverse is
\begin{equation}\label{eq:Lmetrikain}
({}^\star G^{-1})^{ij}=G^{ij}+\frac{\alpha}{1-\alpha a^2}a^i
a^j=G^{ik}(P_T^0+\frac{1}{A}\,P_0)_k{}^j\, .
\end{equation}

The effective metric ${}^\star G^{eff}_{ij}$ has the same form as
in the dilaton free case up to the substitution
$G_{ij}\to\;{}^\star G_{ij}$
\begin{equation}\label{eq:effm}
{}^\star G^{eff}_{ij}={}^\star G_{ij}-4(B\;{}^\star
G^{-1}B)_{ij}=G_{ij}^{eff}-\alpha a_i a_j-\frac{4\alpha}{1-\alpha
a^2}(Ba)_i(aB)_j\, ,
\end{equation}
where $G^{eff}_{ij}=G_{ij}-4B_{ik}G^{kl}B_{lj}$. In terms of the
projectors, we have
\begin{equation}\label{eq:zvezdaeff}
{}^\star G_{ij}^{eff}=(\Pi_T+\tilde A \Pi_0+\mathcal A \Pi_1)_i{}^k
G^{eff}_{kj}\, ,
\end{equation}
where
\begin{equation}\label{eq:a0a1}
\tilde A=1-\alpha \tilde a^2\, ,\quad \mathcal A=\frac{\tilde A}{A}=\frac{1-\alpha \tilde
a^2}{1-\alpha a^2}\, .
\end{equation}
With the help of (\ref{eq:zvezdaeff}), for $\tilde A\neq0$ and $A\neq0$, we obtain
\begin{equation}\label{eq:zvezdaeffin}
({}^\star
G_{eff}^{-1})^{ij}=(G_{eff}^{-1})^{ik}(\Pi_T+\frac{1}{\tilde A}
\Pi_0+\frac{1}{\mathcal A} \Pi_1)_k{}^j=(G_{eff}^{-1})^{ij}+\frac{\alpha}{1-\alpha \tilde a^2}\left[ \tilde a^i \tilde a^j+4(B\tilde a)^i (\tilde a B)^j\right] \, .
\end{equation}

According to Eq.(\ref{eq:Lmetrika}) the determinant of ${}^\star G_{ij}$ is of the form
\begin{equation}\label{eq:detzvezda}
\det\;{}^\star G_{ij}=A\det G_{ij}\, ,
\end{equation}
while the determinant of the effective metric ${}^\star G_{ij}^{eff}$ (\ref{eq:zvezdaeff}) is
\begin{equation}\label{eq:detzvezdaeff}
\det\;{}^\star G_{ij}^{eff}=\tilde A \mathcal A\det G_{ij}^{eff}=\frac{\tilde A^2}{A}\det G_{ij}^{eff}\, .
\end{equation}

For $A=0$, we have $\det\;{}^\star G_{ij}=0$ and the vector $a^i$ is singular for the metric ${}^\star G_{ij}$, what is obvious from 
\begin{equation}
{}^\star G_{ij}a^j=A a_i\, .
\end{equation}

For $\tilde A=0$ and $A\neq0$ the effective metric ${}^\star G_{ij}^{eff}$ is singular. From the relations 
\begin{equation}
{}^\star G_{ij}^{eff}\tilde a^j=\tilde A a_i\, ,\quad  {}^\star G_{ij}^{eff}(\tilde a B)^j=\frac{\tilde A}{A} (aB)_i\, ,
\end{equation}
follows that $\tilde a^i$ and $(\tilde a B)^i$ are singular vectors of the star effective metric.

\section{Separation the center of mass variable}
\setcounter{equation}{0}

We will explain separation the center of mass variable and define the corresponding functions $\theta(x)$ and $\Delta(x)$. Let variable $x^i$ satisfies the Poisson bracket
\begin{equation}
\left\lbrace x^i(\tau,\sigma),x^j(\tau,\overline\sigma)\right\rbrace=2\;{}^\star \Theta^{ij}\theta(\sigma+\overline\sigma)\, ,
\end{equation}
where the function $\theta(x)$ is defined as
\begin{equation}\label{eq:fdelt}
\theta(x)=\left\{\begin{array}{ll}
0 & \textrm{if $x=0$}\\
1/2 & \textrm{if $0<x<2\pi$}\, .\\
1 & \textrm{if $x=2\pi$} \end{array}\right .
\end{equation}
Separating the center of mass variable
\begin{equation}\label{eq:cenmassX}
x^i_{cm}=\frac{1}{\pi}\int_0^\pi d\sigma x^i(\sigma)\, ,\quad
x^i(\sigma)=x^i_{cm}+X^i(\sigma)\, ,
\end{equation}
we obtain
\begin{equation}
\left\lbrace X^i(\tau,\sigma),X^j(\tau,\overline\sigma)\right\rbrace={}^\star \Theta^{ij}\Delta(\sigma+\overline\sigma)\, ,
\end{equation}
where the function $\Delta(x)$ 
\begin{equation}\label{eq:DELTA}
\Delta(x)=2\theta(x)-1=\left\{\begin{array}{ll}
-1 & \textrm{if $x=0$}\\
0 & \textrm{if $0<x<2\pi$}\, ,\\
1 & \textrm{if $x=2\pi$} \end{array}\right .
\end{equation}
is different from zero only on the string endpoints.

The same procedure can be applied to variable ${}^\star F(\sigma)$ with notation
\begin{equation}\label{eq:cenmassF}
{}^\star F(\sigma)={}^\star F_{cm}+{}^\star \mathcal F(\sigma)\, .
\end{equation}

\end{document}